# Gate-electric-field and magnetic-field control of versatile topological phases in a semi-magnetic topological insulator

Ryota Watanabe[1], Ryutaro Yoshimi[2], Kei S. Takahashi[2], Atsushi Tsukazaki[3],

Masashi Kawasaki[1,2], Minoru Kawamura[2]* and Yoshinori Tokura[1,2,4]

[1] *Department of Applied Physics and Quantum Phase Electronics Center (QPEC), University of Tokyo, Bunkyo-ku, Tokyo 113-8656, Japan.*

[2] *RIKEN Center for Emergent Matter Science (CEMS), Wako, Saitama 351-0198, Japan.*

[3] *Institute for Materials Research, Tohoku University, Sendai, Miyagi 980-8577, Japan.*

[4] *Tokyo College, University of Tokyo, Bunkyo-ku, Tokyo 113-8656, Japan.*

\* Correspondence to: minoru@riken.jp

# Abstract

Surface states of a topological insulator demonstrate interesting quantum phenomena, such as the quantum anomalous Hall (QAH) effect and the quantum magnetoelectric effect. Fermi energy tuning plays a role in inducing phase transitions and developing future device functions. Here, we report on controlling the topological phases in a dual-gate field-effect transistor of a semi-magnetic topological insulator heterostructure. The heterostructure consists of magnetized one-surface and non-magnetic other-surface. By tuning the Fermi energy to the energy gap of the magnetized surface, the Hall conductivity $\sigma_{xy}$ becomes close to the half-integer quantized Hall conductivity $e^2/2h$, exemplifying parity anomaly. The dual-gate control enables the band structure alignment to the two quantum Hall states with $\sigma_{xy} = e^2/h$ and 0 under a strong magnetic field. These states are topologically equivalent to the QAH and axion insulator states, respectively. Precise and independent control of the band alignment of the top and bottom surfaces successively induces various topological phase transitions among the QAH, axion insulator, and parity anomaly states in magnetic topological insulators.

A three-dimensional topological insulator (TI) hosts gapped bulk bands and gapless Dirac surface states (SSs) protected by time-reversal symmetry[1,2]. When time-reversal symmetry on the SSs is broken by exchange interaction with the magnetic moments, an exchange gap is formed around the band crossing point (Fig. 1(a) left). When the Fermi energy ($E_F$) is tuned within the exchange gap by applying gate-bias or chemical doping, the quantum anomalous Hall effect (QAHE) emerges[3-5] with the quantized Hall conductivity $\sigma_{xy} = e^2/h$ (Fig. 1(b)). The QAHE has been obtained in thin films of TIs doped with magnetic elements Cr (refs. 4 and 6) or V (ref. 7). When opposite signs of exchange gaps appear for top and bottom SSs of TI thin films, e.g., with the up and down directions of magnetizations or the different signs of exchange coupling, another quantum phase termed axion insulator emerges with $\sigma_{xx} = \sigma_{xy} = 0$ (Fig. 1(c)), where $\sigma_{xx}$ is the longitudinal conductivity[8,9]. The axion insulator state is expected to show the quantized topological magnetoelectric (TME) effect[1]. The axion insulator state has been reported in magnetic TI heterostructures possessing antiparallel magnetizations near the top and bottom SSs[10,11].

By making use of the stacking degree of freedom of the TI heterostructure, we can fabricate semi-magnetic bilayer heterostructures of magnetic and non-magnetic TIs, which host the asymmetric SSs of gapped one-surface and Dirac-like other-surface (Fig. 1(d)). One of the possible topological phases in a semi-magnetic TI under zero magnetic field is a parity anomaly state where the half-integer quantized Hall conductivity, characteristic of a single gapped Dirac cone, is observed[12]. The half-integer quantized Hall conductivity has been demonstrated in various type of semi-magnetic TI heterostructures with the $E_F$ tuning to the exchange gap of the magnetic SS[13]. On the other hand, under a high magnetic field ($B$) applied normal to the film plane, the Dirac-like SS forms the Landau levels as schematically shown in Fig. 1(a) (right). When $E_F$ of the bottom magnetic SS locates within the exchange gap, the Landau level formation of the top Dirac surface drives the semi-magnetic TI to the topological states with the Chern number of $C = 1$ or $0$ depending on the $E_F$ position of the top SS between the $N = 0$ and $N = +1$ or $-1$ (ref. 14), as shown in Figs. 1e and 1f. The former $C = 1$ quantum Hall state (Fig. 1(e)) is viewed as the hybrid of the half-integer quantized anomalous (bottom) and half-integer quantized ordinary (top) Hall states. It is topologically equivalent to the QAH state (Fig. 1(b)). On the other hand, the $C = 0$ state (Fig. 1(f)) is topologically equivalent to the axion insulator (Fig. 1(c)). Thus, the semi-magnetic TI heterostructure is a fascinating platform to host those topological phases. In this study, we explore the topological phases and their mutual transformations in a semi-magnetic TI heterostructure.

In this study, we use a dual-gate field-effect-transistor (FET) device in which a semi-magnetic TI heterostructure is sandwiched by the top and bottom gates. Dual-gate FET devices made of TI thin films in the previous studies[15-17] have demonstrated independent tuning of the carrier density of top surface $n_t$ and of bottom surface $n_b$ by changing the two gate voltages. Such independent carrier-tuning for the top and bottom SSs (Fig. 2(a)) is also expected to expand the controllability of topological phases in a semi-magnetic TI heterostructure. To make a dual-gate device, we employ SrTiO$_3$ substrate because it can serve as the bottom gate dielectric owing to its large dielectric constant at low temperatures. The top gate can be fabricated by depositing AlO$_x$ as the gate dielectric on top of the TI film[14]. In addition to the dual gate tuning, the Bi/Sb ratio is one other critical parameter for controlling the Dirac point position of the surface states[18-20].

To this end, we grew a semi-magnetic heterostructure $(Bi_{0.3}Sb_{0.7})_2Te_3$ (7 nm) / $Cr_{0.2}(Bi_{0.44}Sb_{0.56})_{1.8}Te_3$ (2 nm)/ $(Bi_{0.44}Sb_{0.56})_2Te_3$ (1 nm) on a $SrTiO_3$ substrate (Fig. 2(b)) by molecular beam epitaxy. To obtain atomically flat surface, $SrTiO_3$(111) substrates were soaked in deionized water at 85 °C for 90 minutes and annealed in an oxygen atmosphere in a tube furnace at 930 °C for 3 hours[21,22]. The annealed substrates were transferred into the growth chamber and annealed at 400 °C in vacuum under Te beam flux with a beam equivalent pressure (BEP) $P_{Te} = 1.0 \times 10^{-4}$ Pa. $Cr_x(Bi_{1-y}Sb_y)_{2-x}Te_3$ and $(Bi_{1-y}Sb_y)_2Te_3$ were grown at 200 °C and 260 °C, respectively, with BEPs of $P_{Cr} = 0.3 \times 10^{-6}$ Pa, $P_{Sb} = 3.5 \times 10^{-6}$, $P_{Te} = 1.0 \times 10^{-4}$ Pa and $P_{Bi} = 1.5 \times 10^{-6}$ Pa (for $y = 0.70$) and $2.7 \times 10^{-6}$ Pa (for $y = 0.56$). We define the Cr and Sb composition as $x = 2P_{Cr}/(P_{Cr} + P_{Bi} + P_{Sb})$ and $y = P_{Sb}/(P_{Bi} + P_{Sb})$, respectively. Here the Sb composition in this study was optimized to make $E_F$ locating around Dirac point on the films fabricated on $SrTiO_3$ substrates (See Supplementary Materials Fig. S3). As shown in Fig. 2(a), the magnetic element Cr is doped only in the vicinity of the bottom surface to induce an energy gap at the bottom surface. Note here that the bottom layer, $(Bi_{0.44}Sb_{0.56})_2Te_3$ (1 nm), which is typically applied to improve the crystallinity of the $Cr_{0.2}(Bi_{0.44}Sb_{0.56})_{1.8}Te_3$ (2 nm) layer, does not largely affect the properties of the bottom SS[23]. We annealed the grown heterostructured films at 380 °C for 30 minutes to improve the crystal quality. The films were passivated by a capping layer of $AlO_x$ (3 nm) deposited using atomic layer deposition (ALD).

The film was shaped into a Hall bar by using photolithography and chemical wet etching. The etching solution was mixed acid etchant of $H_2O_2:H_3PO_4:H_2O = 1:1:8$. Contact electrodes of Ti/Au were deposited by an electron beam evaporator. After forming a Hall bar, we deposited a 30-nm-thick AlO$_x$ by the ALD as a dielectric layer of the top gate. Then, the top gate electrode was formed by depositing Ti/Au. The width of the Hall bar was 100 μm, and the distance between the voltage electrodes was 150 μm. The dual gating operation was carried out as depicted in Fig. 2(c). In this study, we explore the possible topological phases in Figs. 1(d)-1(f) and their transitions as a function of top and bottom surface carrier densities $n_t$, $n_b$, and magnetic field.

Electrical transport properties were measured in a Quantum Design Physical Property Measurement System and in a dilution refrigerator by a lock-in technique at a frequency $f$ = 13 Hz with excitation current $I$ = 100 nA and 10 nA, respectively. The devices were mounted on a metal plate which we used as a bottom gate electrode (G$_b$). The excitation current was fed between drain (D) and source (S) electrodes (Fig. 2(c)) and the S electrode was connected to the ground level. The top and bottom gate voltage $V_t$ and $V_b$ were applied between G$_t$-S and G$_b$-S. Figures 2(d) and 2(e) show three-dimensional plots for sheet resistivity $\rho_{xx}$ and sheet Hall resistivity $\rho_{yx}$ as a function of $V_t$ and $V_b$ at temperature $T$ = 0.5 K. Here, a small magnetic field of $B$ = 0.02 T was applied perpendicular to the film plane to align the magnetization direction. In both plots, $V_t$ dependence presents clear peaks, which reflects the large controllability of $E_F$ of top Dirac SS across the Dirac point. In contrast, $V_b$ moderately modulates both $\rho_{yx}$ and $\rho_{xx}$. At the gapped bottom SS, the rather insensitive $V_b$ dependence may come from the exchange-gap formation as previously observed in symmetric modulation-doped magnetic heterostructures[23]. However, the large variation of $\rho_{yx}$ with $V_b$ at the peak region (Fig. 2(e)) clearly indicates that the bottom gate can work well to tune the $E_F$ around the gapped energy levels.

For the further analysis, the values of $V_t$ and $V_b$ are converted to the surface carrier densities $n_t$ and $n_b$ via the procedure as detailed in Supplementary Materials. We estimated $n_t$ and $n_b$ using $V_t$, $V_b$, the ordinary Hall coefficient $R_H$ defined in the region of saturated magnetization $R_H = (\rho_{yx}(6T) - \rho_{yx}(3T))/3T$, and the Hall conductivity $\sigma_{xy} = \rho_{yx}/(\rho_{xx}^2 + \rho_{yx}^2)$ around the charge neutral point (CNP) at $B = 0.02$ T (refs. 15 and 24). Note that the normalized carrier density at CNP is nominally defined as $n_t = n_b = 0$, which represents that $E_F$ locates at the Dirac point on the top SS and at the mid-point of the gap on the bottom SS, respectively. Figures 3(a) and 3(b) exemplify the $n_b$ dependence of $\rho_{yx}$ and $\sigma_{xy}$ at $T = 40$ mK and $B = 0.02$ T for $n_t = 1.0, 0.0$, and $-1.0 \times 10^{16}$ m$^{-2}$. The variation of $\rho_{yx}$ as a function of $n_b$ is largely dependent on the $n_t$ value, i.e., on the $E_F$ position around the Dirac point of the top SS. In contrast, the $\sigma_{xy}$ - $n_b$ curves for the three $n_t$ conditions are consistent with each other as shown in Fig. 3(b). Here, $\sigma_{xy}$ is deduced via the relation that $\sigma_{xy} = \rho_{yx}/(\rho_{xx}^2 + \rho_{yx}^2)$. With decreasing $n_b$ to $0.0 \times 10^{16}$ m$^{-2}$, the values of $\sigma_{xy}$ tend to saturate at values close to $e^2/2h$ regardless of $n_t$. This behavior is in accord with the previously reported parity anomaly state[13]. In the present study, the $E_F$ position was tuned by electric gating while it was tuned by the Bi/Sb ration in the previous study. As clearly seen in Fig. 3(c), the red region displaying $e^2/2h$ lies nearly parallel contour lines indicating that $\sigma_{xy}$ dominantly depends on $n_b$ (gapped bottom SS), but not on $n_t$. The respective band structures of the top and bottom SSs at the representative points A, B, and C in Fig. 3(c) are schematically depicted in Fig. 3(d). The $(n_t, n_b)$ dependence of $\sigma_{xy}$ indicates that $\sigma_{xy}$ takes values close to $e^2/2h$ irrespective of $n_t$ when the $E_F$ position of the bottom surface locates in the exchange gap of the bottom SS. The observation in Fig. 3 confirms that the parity anomaly state can appear in a fairly wide range of the $E_F$ position of top Dirac SS. The variation of $\sigma_{xy}$ against $(n_t, n_b)$ is consistent with the scenario of the parity anomaly state depicted in Fig. 1(d). We speculate that the observed deviation from $\sigma_{xy}$ from the exact half-integer quantized value $e^2/2h$ is due perhaps to the imperfect suppression of the *bulk* carrier density[14] inside near the magnetized bottom layer of the semi-magnetic structure adjacent to the SrTiO$_3$ gate dielectric or the disorder in the gapped bottom SS.

In addition to the parity anomaly state at zero magnetic field, fascinating quantum states appear as the incipient quantum Hall (QH) state with $C = 1$ as well as the axion insulator state with $C = 0$ by applying a magnetic field $B$ normal to the film plane, as predicted in Figs. 1(e) and 1(f). Figures 4(a) and 4(b) show $n_t$ dependence of $\sigma_{xx}$ and $\sigma_{xy}$ for $B$ = 0.02, 0.2, 3, 6, 10, and 14 T at $T$ = 40 mK. $E_F$ of the bottom SS is tuned into the gap, corresponding to $n_b = 0.0 \times 10^{16}$ m$^{-2}$. Two minima in $\sigma_{xx}$ develop with increasing $B$ at around $n_t = -0.5 \times 10^{16}$ and $0.8 \times 10^{16}$ m$^{-2}$. At $n_t = -0.5 \times 10^{16}$, $\sigma_{xy}$ approaches $e^2/h$ with increasing $B$ while $\sigma_{xy}$ approaches 0 at $n_t = 0.8 \times 10^{16}$ m$^{-2}$, as indicated by thick vertical arrows in Fig. 4(b) (Magnetic field dependence of conductivities for the $C = 1$ and $C = 0$ states are respectively shown in Figs. S2(a) and S2(b) of Supplementary Materials). Also, at these carrier densities, the slopes $\sigma_{xy}$ - $n_t$ curves tend to be flat as increasing magnetic fields, showing the incipient plateaus in $\sigma_{xy}$. The simultaneous observation of the reduction in $\sigma_{xx}$ and the incipient plateaus close to the quantized values of $\sigma_{xy}$ ($e^2/h$ and 0) suggests the development of topological quantum states as assigned to the quantum Hall insulator (Fig. 1(e)) and the axion insulator (Fig. 1(f)), respectively. Even though the $E_F$ position is well controlled, finite $\sigma_{xx}$ remains at the quantized condition. The origin of the remained $\sigma_{xx}$ is probably attributed to the hopping conduction in the SS or the residual bulk contribution. To suppress $\sigma_{xx}$, the number of impurities in the heterostructure must be reduced by further optimizing of the growth temperature, the growth rate, and the pre-growth treatment of the SrTiO$_3$ substrate.

In addition to the $C = 0$ and $C = 1$ states, an invariant point in $\sigma_{xx}$ with respect to $B$ is observed at $n_t \sim 0.0 \times 10^{16}$ m$^{-2}$, where $\sigma_{xy}$ stays roughly at $e^2/2h$ with increasing $B$. With increasing magnetic field, the half-integer quantized Hall conductivity, characteristic of parity anomaly, can persist only in the region near the CNP, and $\sigma_{xy}$ evolves to either 0 or $e^2/h$ in the other area. In color maps for $\sigma_{xx}$ and $\sigma_{xy}$ as functions of $n_b$ and $n_t$ at $B = 14$ T and $T = 40$ mK shown in Figs. 4(c) and 4(d), these quantum states $C = 1$ ($\sigma_{xy} = e^2/h$) and $C = 0$ ($\sigma_{xy} = 0$) are clearly observed and well separated by the vertical line around $n_t \sim 0.0 \times 10^{16}$ m$^{-2}$. The topological phase transition between the $C = 1$ and $C = 0$ states is induced by tuning $n_t$ while these states are widely stable against variation of $n_b$, in accord with the results shown in Fig. 3. By the variation of $n_t$, the occupancy of Landau levels in the top Dirac SS could be systematically tuned. As schematically shown in Figs. 4e and 4f, the Landau levels with $N =$ integer (black lines) are formed by the application of $B$. The $E_F$ position (red dashed lines) on the top SS effectively shifts with $n_t$ from in between $N = -1$ and 0 to in between $N=0$ and at 1, promoting the change from $\sigma_{xy} = e^2/h$ (Fig. 4(e)) to $\sigma_{xy} = 0$ (Fig. 4(f)). Here, it is to be noted that this transition cannot occur in a trivial insulator. The transition driven by change of $n_t$ confirms that the observed $\sigma_{xy} = 0$ is attributed to the axion insulator as depicted in Fig. 1(f), discriminating the possibility of a trivial insulator with large $\rho_{xx}$. Any other phase transitions were not observed by control of $n_b$ in the present sample, while these QH states were extinguished by locating $E_F$ away from the exchange gap. The Zeeman effect may also can contribute to changing the energy gap. The Zeeman effect should appear most remarkably in the energy shift of the zeroth Landau level.

However, as shown in Fig. 4(a), the $\sigma_{xx}$ peak corresponding to the zeroth Landau level did not shift clearly in the present study. We speculate that the disorder-induced broadening of the $\sigma_{xx}$ peak masks the energy shifts by the Zeeman effect.

Another point to be noted is that the QH state with $\sigma_{xy} = 0$ could be a possible platform for exploring the TME effect instead of the conventional axion insulator shown in Fig. 1(c). In the present $C = 0$ state, the top and bottom SSs are respectively gapped due to the Landau level formation with magnetic field and the exchange field, and hence the Hall conductivity of the two SSs is also cancelled out with opposite sign in the Berry curvature. The present $C = 0$ QH state with $\sigma_{xy} = 0$ satisfies the above requirements for the TME effect as well as the known axion insulators (Fig. 1(f)). Robustness of the $C = 0$ state against strong magnetic field could expand experimental options[25,26], because the axion insulators stabilized by the antiparallel magnetizations can only survive in a narrow range of external magnetic field due to small difference of coercive field of the magnetic layers.

To summarize, we have investigated the dual-gate control for topological phase transitions in a FET made of a semi-magnetic TI $Cr_x(Bi_{1-y}Sb_y)_{2-x}Te_3$ / $(Bi_{1-y}Sb_y)_2Te_3$ heterostructure. We observed the robust half-integer quantized Hall conductivity with $E_F$ of the bottom SS locating in the exchange gap, regardless of the $E_F$ position of the top Dirac SS. Our study further demonstrated a systematic control of the quantum phase transition between the $C = 1$ quantum Hall state with $e^2/h$ and the $C = 0$ axion insulator state by applying $B$. The observed $C = 0$ state with $\sigma_{xy} = 0$ is topologically equivalent to the axion insulator of the exchange interaction origin and is expected to show the quantum magnetoelectric effect in a more robust magnetic-field range compared to the axion insulator of the exchange interaction origin. The dual-gate device based on semi-magnetic TI heterostructures will pave a way to access to intriguing quantum topological phases and their spintronic and photonic functions of the topological surface states.

**Acknowledgements**

The authors thank Tian Liang for the collaboration in designing the dual-gate device structure. This work was partly supported by the Japan Society for the Promotion of Science through JSPS/MEXT Grant-in-Aid for Scientific Research (Nos. 22J13360, 18H01155, and 22H04958).

**Competing interests**

The authors declare no competing interests.

**Data Availability**

The data that support the findings of this study are available from the corresponding author upon reasonable request.

**Additional information**

Supplementary Materials are available for this paper.

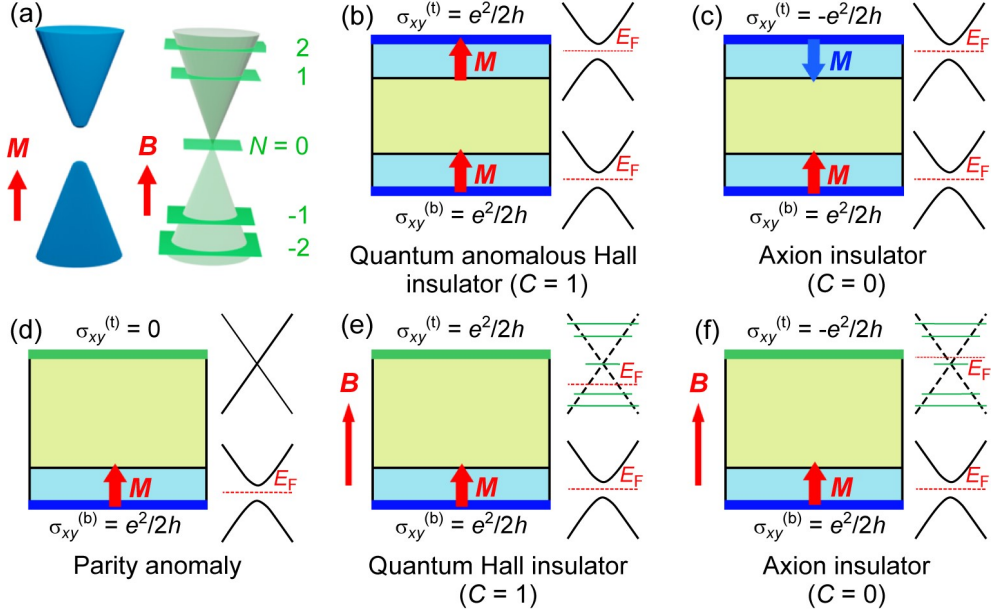

**Fig. 1.** (a) Left: the gapped surface state (SS) induced by the exchange interaction with perpendicular component of magnetization (*M*), and right: the Landau level formation in the Dirac SS by applying magnetic field (*B*). (b) Quantum anomalous Hall insulator (Chern number *C* = 1) with the parallel top and bottom *M*. (c) Axion insulator (*C* = 0) with the antiparallel *M*. In (b) and (c), the left panels show cross-sectional TI trilayers heterostructure with the non-magnetic TI layer sandwiched by two magnetic layers of top and bottom, while the right panels show the top and bottom SS and the Fermi level ($E_F$) position. (d) The semi-magnetic TI with only one-side (bottom) magnetized surface and hence gapped SS. (e)(f) The quantum Hall insulator (*C* = 1) state (e) and the axion insulator (*C* = 0) state (f) as derived from the semi-magnetic TI by application of external magnetic field (*B*). For all these topological states shown here, the $E_F$ is assumed to locate in the gap caused by the exchange field and/or between the Landau levels with index *N* = 0 and *N* = ±1.

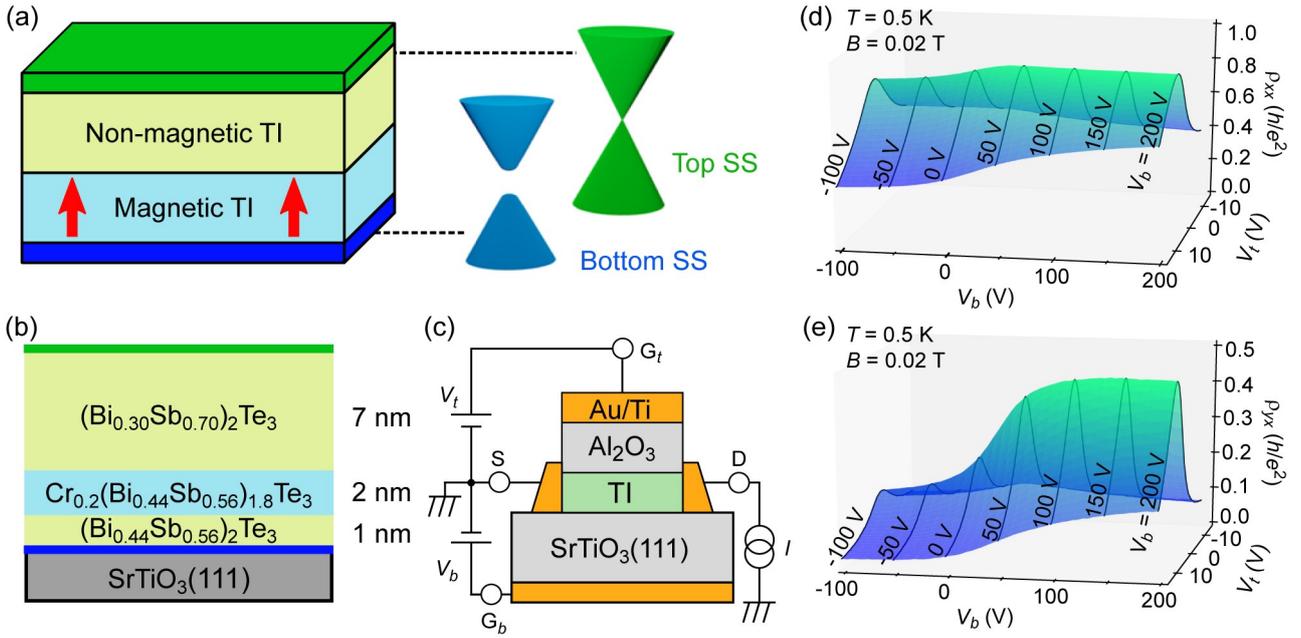

**Fig. 2.** (a) A schematic of a semi-magnetic topological insulator (TI) with asymmetric surface states. (b) Cross-sectional schematic of a semi-magnetic TI heterostructure on a SrTiO$_3$ substrate. (c) Cross-sectional device structure and measurement setup of a dual-gate field-effect transistor (FET). (d)(e) The top gate voltage $V_t$ and bottom gate voltage $V_b$ dependence of $\rho_{xx}$ (d) and $\rho_{yx}$ (e) at $T$ = 0.5 K and $B$ = 0.02 T.

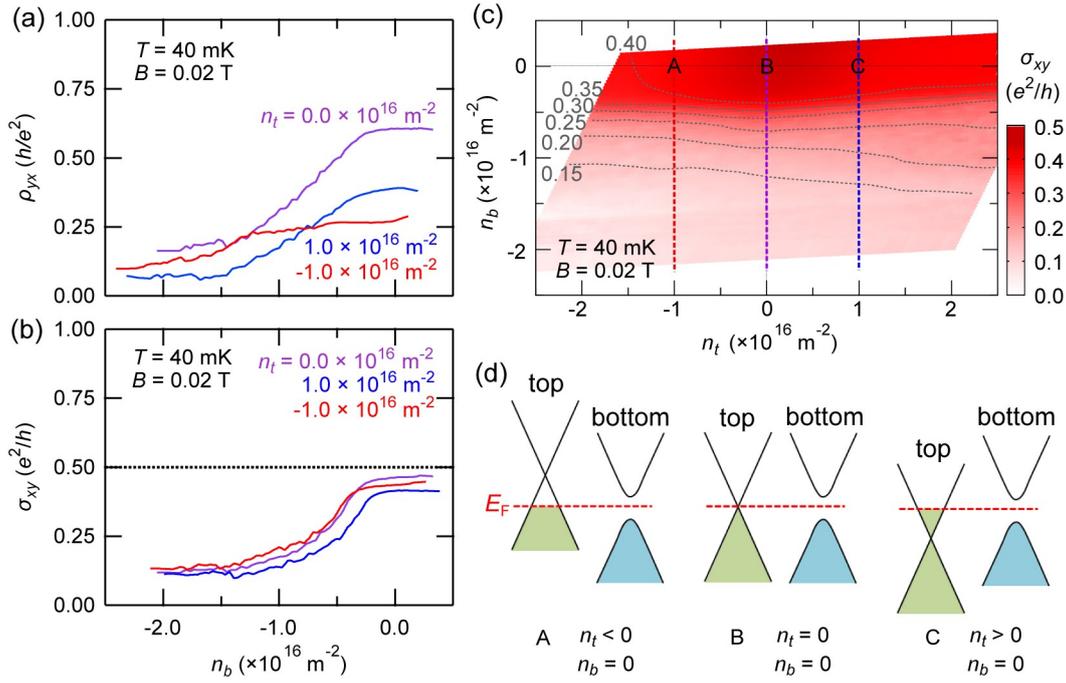

**Fig. 3.** (a)(b) Bottom surface carrier density $n_b$ dependence of $\rho_{yx}$ (a) and $\sigma_{xy}$ (b) under the carrier density of top SS is fixed at $n_t$ = -1.0, 0.0 and 1.0 × 10$^{16}$ m$^{-2}$ (shown as vertical broken lines in Fig. 2(c)) at $T$ = 40 mK and $B$ = 0.02 T. (c) A contour plot for $\sigma_{xy}$ as functions of $n_t$ and $n_b$. (d) Band diagrams for the ($n_t$, $n_b$) points with labels A, B and C in the panel (c) with the relative positions of the $E_F$ (horizontal red dashed red lines).

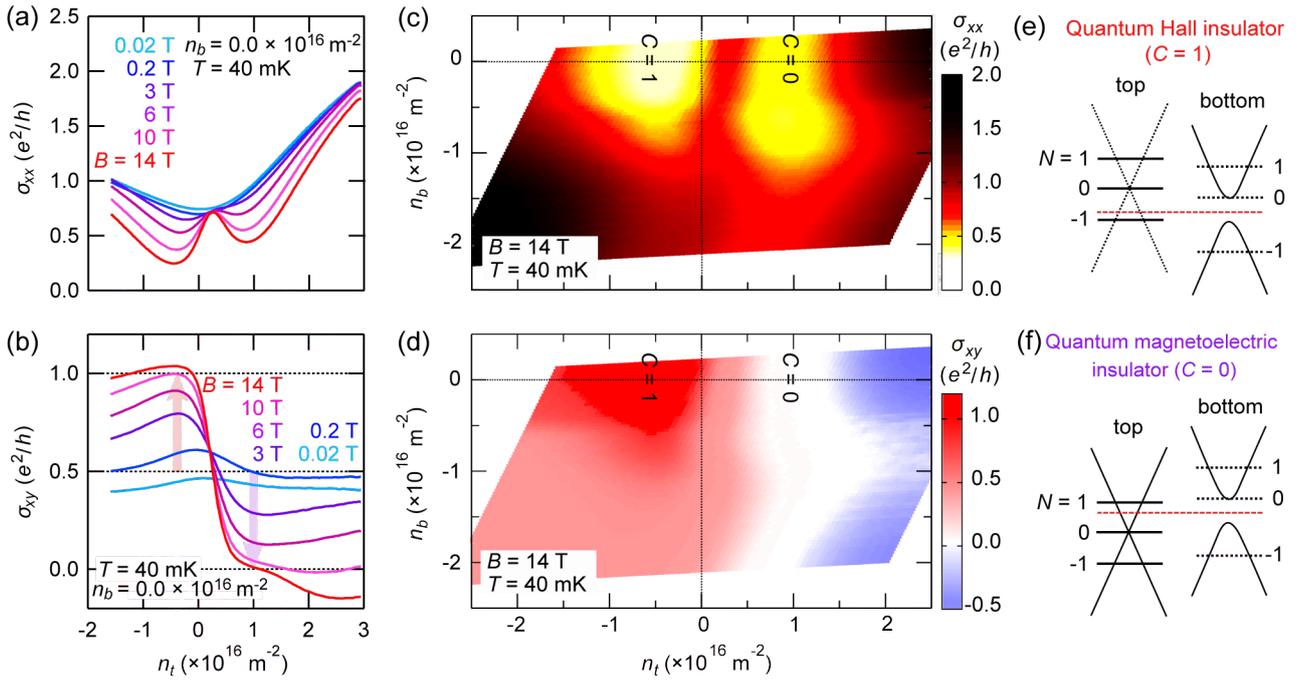

**Fig. 4.** (a)(b) Top surface carrier density $n_t$ dependence of $\sigma_{xx}$ (a) and $\sigma_{xy}$ (b) for $n_b = 0.0 \times 10^{16}$ m$^{-2}$ at $B$ = 0.02, 0.2, 3, 6, 10 and 14 T and $T$ = 40 mK. Here, $n_t$ and $n_b$ represent the carrier density of the top and bottom surface state, respectively. (c)(d) Contour maps for $\sigma_{xx}$ (c) and $\sigma_{xy}$ (d) as a function of $n_t$ and $n_b$ at $B$ = 14 T and $T$ = 40 mK. (e)(f) Band diagrams are shown for (e) the $C$ = 1 state and (f) the $C$ = 0 state as labeled in the panels (c) and (d).